\documentclass{appolb}
\usepackage{graphicx}
\usepackage[]{hyperref}
\hypersetup{
	pdfborder={0,0,0},
	breaklinks=true,
	pdftitle=PDF dependence on parameter fits from hadronic data,
	pdfauthor=Zahari Kassabov
}
\newcommand{\as}{\alpha_S}
\begin{document}
\title{PDF dependence on parameter fits from hadronic data%
\thanks{Presented at The Final HiggsTools Meeting, 11-15 September
2017, IPPP}%
}
\author{Zahari Kassabov
\address{Cavendish Laboratory, University of Cambridge, Cambridge CB3
0HE, United Kingdom} \\
}
\maketitle
\begin{abstract}
We present a discussion on the methods for extracting a given parameter
from measurements of hadronic data, with particular focus on
determinations of the strong coupling constant. We show that when the
PDF dependency on the determination is adequately taken into account, the
dispersion between the results from different measurements is
significantly reduced. We speculatively propose the concept of
\emph{preferred value} of a parameter from a particular dataset.
\end{abstract}

CAVENDISH-HEP-18-04

\section{Introduction}
Since the beginning of its operation, the Large
Hadron Collider (LHC) has produced a wealth of experimental data,
which has been used most notably to establish the existence of the
Higgs Boson~\cite{Aad:2012tfa,Chatrchyan:2012xdj}. These results have
not only been used to validate the Standard Model but also to
accurately measure its parameters. This involves matching experimental
data to theoretical predictions.
The experimental results are typically obtained for
hadronic cross sections ($\sigma_{pp\to X}$) for a given final state
$X$, while the theory predictions are usually computed for hard
(partonic) quantities in the framework of Perturbation Theory,
$\hat{\sigma}_{ab\to X}$. The two quantities can be related by universal
parton distributions~\cite{Ellis:1991qj}: Using the notation from
Ref.~\cite{Forte:2010dt}, we have
\begin{equation}
	\label{eq:hadroproduction} \sigma_{pp\to X}(s, M_X^2)
	= \sum_{a,b}\int_{x_{\min}}dx_1dx_2 f_a(x_1, M_X^2) f_b(x_2,
M_X^2) \hat{\sigma}_{ab\to X}(x_1x_2s, M_X^2) \ .
\end{equation}
The
Parton Distribution Functions (PDFs) of the proton, $f_a(x,Q^2)$
cannot be computed from first principles and instead need to be
determined from experimental data. These PDFs are themselves also
obtained by appropriately matching experimental data (usually from
a wide variety of physical processes) to the predictions for the
corresponding partonic cross sections. Thus, a PDF fitting methodology
can be viewed as an algorithm that takes as input a set of
experimental measurements, together with a set of theory assumptions,
and produces a set of parton distribution functions, with an estimate
of their uncertainties. Roughly speaking, the PDFs are obtained by
minimizing a $\chi^2$ error function
\begin{equation}
\label{eq:chi2}
	\chi^{2}\left[\{\theta\},\{\alpha\},\mathcal{D}\right]=
	\sum_{I,J=1}^{N_{\mathcal{D}}}
	\left(T_{I}[\{\theta\},\{\alpha\}]-D_{I}\right) C_{IJ}^{-1}
\left(T_{J}[\{\theta\},\{\alpha\}]-D_{J}\right) \ , \end{equation}
where $\{\theta\}$ represents the set of parameters that determine the
PDF functional form (e.g. the neural network parameters in the case of
the NNPDF\cite{Ball:2017nwa} parametrization), $\{\alpha\}$ is the set of theoretical
parameters used as input (such as the value of the strong coupling
constant $\as(M_Z)$ or the masses of the heavy quarks),
$\mathcal{D}=\{D_1\dots D_{N_\mathcal{D}}\}$ is the set of input
experimental data points (which it will be useful to call
\emph{global} dataset), $T_I$ are the theory predictions corresponding
to the data $\mathcal{D_I}$, and $C_{IJ}$ is experimental covariance
matrix. More realistically, a state of art PDF determination
incorporates schemes to compensate~\cite{Ball:2009qv} biases due to
normalization uncertainties~\cite{D'Agostini:1993uj}, regularization
mechanisms to avoid overfitting (for example, cross
validation~\cite{Ball:2014uwa}) and self validating strategies such
closure tests~\cite{Ball:2014uwa} or
tolerances~\cite{Dulat:2015mca,Harland-Lang:2014zoa}. Usually  the
theory parameters $\{\alpha\}$ are kept fixed while the PDF parameters
$\{\theta\}$ are optimized. However it is in principle
possible to simultaneously optimize for some theory parameter, such as
the $|V_{cs}|$ of the CKM matrix~\cite{Ball:2009mk}.

We recall that any theoretical prediction for hadronic observables
$\sigma_X$, depends trough Eq.~\ref{eq:hadroproduction} both on the
choice of PDF fitting methodology and on the data used as input for
the PDF fit (even though they are often provided together in the form
of a set of PDFs, e.g. as grids in the LHAPDF~\cite{Buckley:2014ana}
format).

\section{Parameter determination from hadronic data}

We now turn our attention to the formalism used to extract theory
parameters from the best fits to hadronic data. While the discussion
applies in general, we restrict ourselves to the determination of the
strong coupling constant, $\as$. The value of $\as$ is generally
quoted at the mass of the $Z$ boson and is
usually~\cite{Butterworth:2015oua,deFlorian:2016spz} taken to be
consistent with the \emph{World Average} produced by the Particle Data
Group~\cite{Patrignani:2016xqp}. Two \emph{categories} of
determinations based on hadronic measurements enter the World Average:
Those based on
PDFs~\cite{alekhin2012parton,jimenez2009dynamical,harland2015uncertainties,ball2012precision},
which are essentially obtained by optimizing Eq~\ref{eq:chi2} as
a function of $\as$, and the $t\bar{t}$ production, currently
including
only the CMS measurement at
7 TeV~\cite{Chatrchyan:2013haa}, which is instead based on minimizing
over a $\chi^2$ function that considers explicitly the $t\bar{t}$
data only (we call this the \emph{partial $\chi^2$ method}). We
shall discuss the relation between these categories, and also try to
elucidate the noticeable fact that determinations of $\as$  based on
an hadronic dataset, such as Ref~\cite{Chatrchyan:2013haa} as well
as more recent ones like Ref~\cite{Andreev:2017vxu}, give
significatively different results from the determinations based on
the PDFs that they use as input to compute the predictions in
Eq.~\ref{eq:hadroproduction}.

\subsection{The \emph{Partial $\chi^2$ method}}
\label{sec:partialchi2}

Several recent determinations of $\as$ based on hadronic
data~\cite{Chatrchyan:2013haa,Bouzid:2017uak,Aaboud:2017fml,Klijnsma:2017eqp,Andreev:2017vxu,Johnson:2017ttl}
implement the following procedure, which we shall dub \emph{Partial
$\chi^2$}:

\begin{enumerate}
\item Consider some experimental measurement of hadronic data,
$\mathcal{P}$. For example, $t\bar{t}$
production~\cite{Chatrchyan:2013haa,Klijnsma:2017eqp}, Prompt photon
events~\cite{Bouzid:2017uak} jet production~\cite{Aaboud:2017fml,
Andreev:2017vxu}, and $Z$+jet production~\cite{Johnson:2017ttl}.
\item Compute theory predictions at discrete values of $\as$,
following Eq.~\ref{eq:hadroproduction} and suitably interpolating the
results from PDF sets fitted with different values of $\as$
(i.e. where $\as(M_Z)$ is a fixed parameter in
Eq.~\ref{eq:chi2}).  \item Construct a profile $\chi_{\mathcal{P}}^{2}(\as)$ characterizing
the agreement between data and theory: Analogously to
Eq.~\ref{eq:chi2} we have
\begin{equation}
\label{eq:partialchi2}
\chi^{2}_p\left[\alpha_{S},\mathcal{P}\right]=
\sum_{I,J=1}^{N_{\mathcal{P}}}\left(
T_{I}[\alpha_{S}]-D_{I}\right)
C_{IJ}^{-1}
\left(T_{J}[\alpha_{S}]-D_{J}\right)
,
\end{equation}
where now the sum is over the partial dataset $\mathcal{P}$.
\item Determine the best fit value of $\alpha_{S}$ as the minimum of the
profile.
\end{enumerate}

We point out that the recommendation~\cite{Butterworth:2015oua} for
estimating $\as$ uncertainties on the PDFs, namely obtaining the final
result with an upper and a lower PDF variation of $\as(M_Z)$ does not
apply when fitting $\as$ itself. In this case the value of $\as$
should be kept matched with the rest of the calculation. Note that
this does not imply that theory  parameters cannot be fixed in PDF
fits by default: For example the value of $\as$ itself is fixed in the
PDF4LHC recommendation~\cite{Butterworth:2015oua} to a value
consistent with the PDG average~\cite{Patrignani:2016xqp} on the
grounds that it takes into account more information than that provided
by hadronic data; we may trade some internal consistency of the input
$\mathcal{D}$ within the PDF fitting framework with potentially more
reliable external constrains on the theory parameters. On the other
hand, theoretical parameters  that are to be fitted do certainly have
to be varied consistently in the PDFs. This is a required condition,
but, as we argue next, not sufficient.

We now discuss the relation between the partial $\chi^2$ method we
just described and the dataset used to fit the PDF by optimizing the
\emph{global} $\chi^2$, Eq.~\ref{eq:chi2}.  In particular it is
pertinent to examine why does the partial $\chi^2$ appear to constrain
$\as$ in all the examples above. That is, why is the value of
$\chi^{2}_p\left[\alpha_{S},\mathcal{P}\right]$ different at different
values of $\as$?

\subsection{PDF and $\as$ determination from a partial dataset}
\label{sec:thatprocessonly}

We note that if the only data used to fit the PDFs was any of the
partial datasets above (such as e.g. $t\bar{t}$ production), so that
$\mathcal{D}=\mathcal{P}$ then  we would certainly not have enough
constraints to determine the PDFs and $\as$ simultaneously: In fact,
we would be able to obtain an adequate fit, characterized by
$\chi^2/(N_\mathcal{D}-1)\approx1$ (see Ref~\cite{Ball:2014uwa} for an
extended discussion) for any reasonable value of $\as$. We would
however have big PDF uncertainties, associated to the kinematic
regions that are not constrained by $\mathcal{P}$.  For example if we
were to fit PDFs to $t\bar{t}$ production data only, we could obtain
a good fit at a higher value of $\as(M_Z)$ by compensating it with
a reduced gluon momentum fraction large $x$ as we will show next in
a more general situation. Therefore for $\mathcal{D}=\mathcal{P}$, the
partial $\chi^2$ in Eq.~\ref{eq:partialchi2}  is flat and does not
allow to determine $\as$ (in this case, $\chi^2_\mathcal{P}$ is also
the global $\chi^2$, Eq.~\ref{eq:chi2}).

It follows that for these relatively small datasets, the
$\chi^{2}_p\left[\alpha_{S},\mathcal{P}\right]$ profile fundamentally
measures the disagreement between the partial data set $\mathcal{P}$
and the dataset included in the PDF fit, $\mathcal{D}$, as a function
of $\as$.

\subsection{Inconsistency of the partial $\chi^2$ method}

\begin{figure}[htb]
\centerline{%
\includegraphics[width=12.5cm]{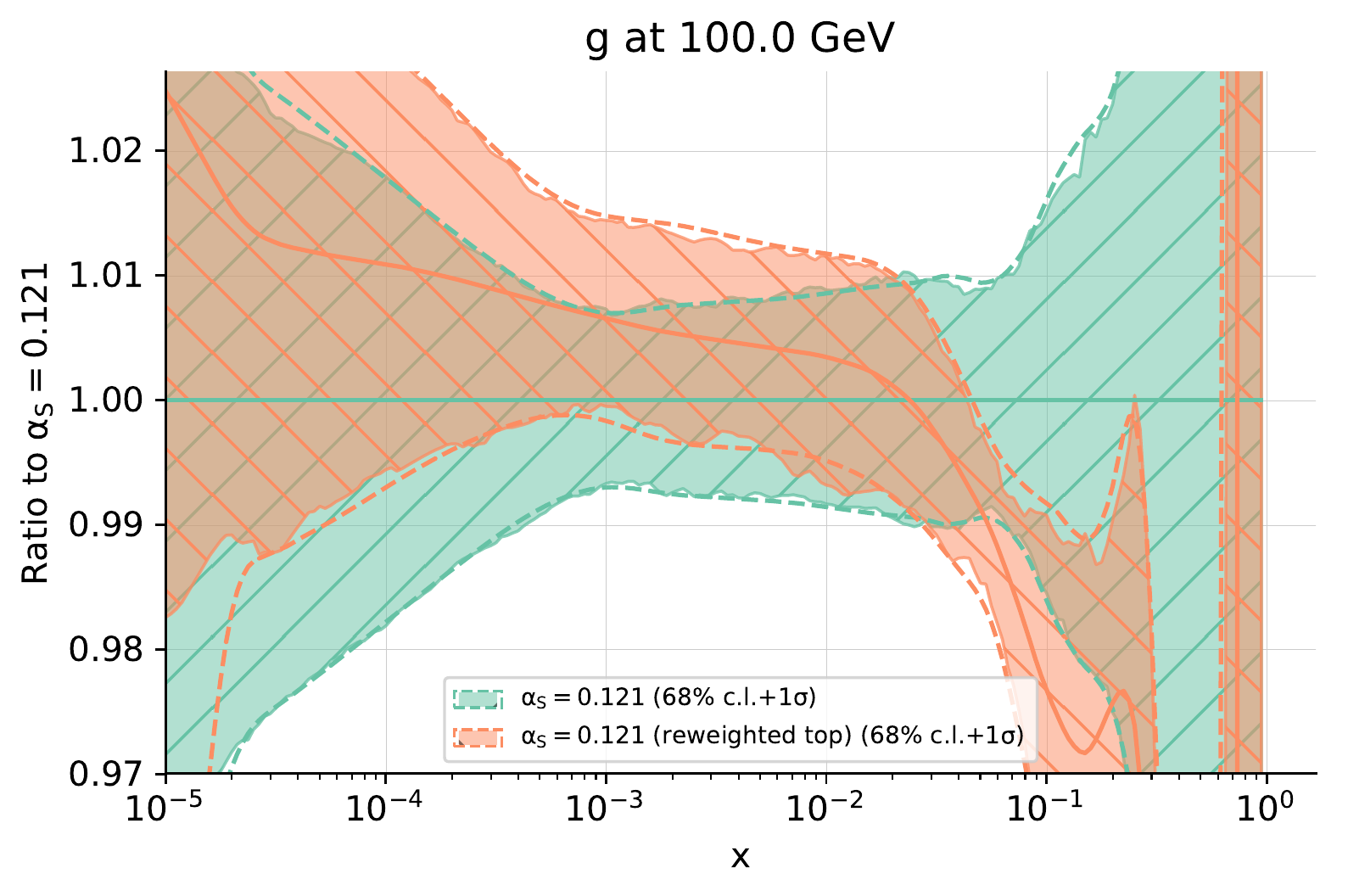}}
\caption{Comparison of gluon PDF between an NNLO-like global fit at NNLO
where we have set $\as(M_Z)=0.121$ and a fit with the only difference
that the  weight with which the $t\bar{t}$ production data enters the
fit has been multiplied
by 15. The reweighting causes noticeable decrease in the gluon at
	large $x$ (but yet roughly within uncertainties) to accommodate the
$t\bar{t}$ data which is, which is then described optimally, with
$\chi^2_{t\bar{t}}/d.o.f.=1.02$, to be compared to
$\chi^2_{t\bar{t}}/d.o.f.=1.42$ before the reweighing. The improvement
of the description of the $t\bar{t}$ data comes at the cost of
a deterioration in the global $\chi^2$ ($\chi^2/d.o.f=1.215$
before the reweighing and $\chi^2/d.o.f=1.229$ afterwards).
}
\label{fig:supertop}
\end{figure}

The partial $\chi^2$ method neglects the fact that the dataset used in
the PDF fits, $\mathcal{D}$,  constrains $\as$ itself, i.e. that the
minimum of Eq.~\ref{eq:chi2} adopts significantly different values for
different values of $\as$. That is, given the measurement
$\mathcal{P}$, if one makes enough assumptions on the input data of
the PDFs to be able to extract $\as(M_Z)$ with competitive
uncertainties, then the prior over $\as$ is not uniform. One cannot
simply disregard the constrains from $\mathcal{D}$ on the theory
parameters $\{\alpha\}$ while utilizing them for the PDF parameters
$\{\theta\}$.  In particular, this can lead to evident inconsistencies
such as the value selected by the partial $\chi^2$ method being
excluded by the PDF on which the theoretical prediction
Eq~\ref{eq:hadroproduction} is based. This is then a logical
contradiction, because the result, which, as we have shown in
Sec.~\ref{sec:thatprocessonly}, is based on the agreement with
$\mathcal{D}$, is grounded on a prior that is internally inconsistent
to begin with.
Moreover, the best fit PDFs away from the global minimum in
$(\{\alpha\}, \{\theta\})$ are subject to a large degree of
arbitrariness: in an ideal PDF fit where all theory and data are
correct, every dataset has
a $\chi^2$ per degree of freedom, $\chi^2/d.o.f\approx 1$. The
$\chi^2$ increases when instead not all the data can be accommodated (e.g. because
the \emph{wrong} value of $\as$ has been given as input). In this
case, the result of
the fit depends on the number of points belonging to each particular
dataset, in such a way that the smaller a  dataset is (in comparison to others
which cannot be fitted simultaneously), the less advantageous is it
for the global figure of merit Eq.~\ref{eq:chi2} to bend the PDF in
order to accommodate it.  This is clear in the case of the $t\bar{t}$
data in the NNPDF 3.1~\cite{Ball:2017nwa} fits. The default dataset
includes a total of 26 $t\bar{t}$ production datapoints corresponding
to the ATLAS~\cite{Aad:2014kva,Aaboud:2016pbd,Aad:2015mbv} and
CMS~\cite{Khachatryan:2016mqs,CMS:2016syx,Khachatryan:2015oqa}
measurements of the total cross sections and differential
distributions, computed at
NNLO~\cite{Czakon:2015owf,Czakon:2016dgf}(see Ref~\cite{Ball:2017nwa}
for details). The $t\bar{t}$ data has a large sensitivity to $\as$ but
a low statistical weight in the fit (26 points to be compared to
3979 in total). Therefore its description (i.e. the partial $\chi^2$,
Eq.~\ref{eq:partialchi2}) deteriorates rapidly as we move move $\as$
away from the best fit value. However, we can modify the assumptions on
$\mathcal{D}$ insisting that the $t\bar{t}$ data is described at
any value of $\as$. For example we set $\as(M_Z)=0.121$ where the top
data is not so well described in a default NNPDF fit that optimizes
Eq~\ref{eq:chi2} on a large dataset (we have
$\chi^2_{t\bar{t}}/d.o.f.=1.42$) and increase the statistical weight
of the top data by fitting 15 identical copies of it. The effect of
the reweighting is to
greatly improve the description of $t\bar{t}$ (the partial $\chi^2$
becomes $\chi^2_{t\bar{t}}/d.o.f.=1.02$) while slightly
deteriorating the global $\chi^2$. The most significative change
between the default fit and that with increased weight
happens in the gluon PDF, which is nevertheless compatible within PDF
uncertainties, as we show in Fig.~\ref{fig:supertop}. Indeed, because
of the high degeneracy in the space of PDF parameters, $\{\theta\}$,
important variations in the input assumptions (that e.g. change
drastically the partial $\chi^2$) can be reabsorbed into relatively
small changes in the PDFs (both in terms of deterioration of the
global $\chi^2$ and distances in PDF space). In this way we have
demonstrated that the partial $\chi^2$ does not measure significant
physical properties of the hard cross section, but rather properties
of the PDF minimization.

In summary, we propose that the most statistically rigorous way to
produce an $\as$ determination from the measurement $\mathcal{P}$ is
to include it in a PDF fit and determine simultaneously $\as$ and the
PDFs based on the global $\chi^2$ that now includes $\mathcal{P}$ as
well as the rest of the data $\mathcal{D}$. Therefore if $\mathcal{P}$
was already included in the $\mathcal{D}$ the result from optimizing
the global $\chi^2$ profile Eq~\ref{eq:chi2} would be unchanged. Since
there is no way to disentangle the $\mathcal{D}$ dependence from
Eq.~\ref{eq:hadroproduction}, this method  is no more PDF dependent
than the partial $\chi^2$ minimization, but it solves the shortcomings
that we have described.  The correction on the value of $\as$ when
$\mathcal{P}$ is included will either be a small or point to a flaw in
the theory, experiment description or fitting methodology.  An
important advantage is that the process will then be treated using the
full fledged PDF fitting machinery (as opposed to a naive minimization
of Eq.~\ref{eq:chi2}). In particular, this takes care of implementing
the correct treatment of the normalization uncertainties, which has
been observed to make a significant difference in an $\as$
determination~\cite{31as}. We conclude that it is questionable to
consider hadronic results as independent constrains on $\alpha_S$ in
World averages, rather than as corrections to the results from the
prior PDFs.

\section{Preferred values}

While we have concluded that the quantities suitable for inclusion in
global averages are those based on minimizing the global $\chi^2$
profile, Eq.~\ref{eq:chi2}, it is nevertheless interesting to define
a \emph{preferred $\as$} value from a given dataset $\mathcal{P}$.
Some possible usages include the assessment of the constraints
provided by the measurement, and  possibly the study of the higher
order corrections (e.g. one could take the dispersion over ensemble of
preferred values of a suitable set of processes as an estimate of
Missing Higher Order Uncertainty).  We first list some desirable
properties that such definition should have.
\begin{itemize}
\item Independent on the relation between the number of points in the
dataset of interest, $N_\mathcal{P}$ and those in the global
dataset, $N_{\mathcal{D}}$. Clearly, if we are interested
in intrinsic physical properties, the number of points in the
dataset should not change the result.
\item Explicitly depend on the global dataset used in the PDF fit
$\mathcal{D}$. Since, as discussed in the previous section, in general
we cannot get rid of the dependence on $\mathcal{D}$, it needs to be
clearly acknowledged.
\item Converge to the determination from $\mathcal{P}$ alone,
in the sense described in Sec.~\ref{sec:thatprocessonly} when it determines $\as$ by itself.
While this definition is likely more interesting for smaller,
experimentally cleaner, datasets, this is a logical asymptotic
property.
\end{itemize}
The partial $\chi^2$ method discussed in
Sec.~\ref{sec:partialchi2} has none of these properties
and therefore it is not a particularly good definition of preferred
value (it may however approximate the third property reasonably well
in practice). On the other hand, the exercise illustrated in
Fig~\ref{fig:supertop} points at a definition that satisfies them:

\begin{description}
\item[Preferred value of $\as$ for the data $\mathcal{P}$] The value
of $\as$ that corresponds to the minimum of the global $\chi^2$ over
values of $\as$ and PDF parameters $\{\theta\}$, when
the  PDF parameters are restricted to result in a \emph{good fit} for
$\mathcal{P}$ within its experimental uncertainties, for all values of
$\as$.
\end{description}

The value is preferred in the sense that the constraints from
$\mathcal{P}$ take precedence over those from $\mathcal{D}$, in
particular regardless of the number of points, thereby satisfying the
first requirement. Once the constrains from $\mathcal{P}$ are
enforced, a global $\chi^2$ which includes $\mathcal{D}$ is minimized,
thus satisfying the second condition.

The main difficulty is to algorithmically specify what a \emph{good
fit} means: Intuitively if the dataset if self consistent at a given
value of $\as$, then we require that $\chi_\mathcal{P}^2/d.o.f\approx 1$. If this
is the case at every relevant value of $\as$ then the partial
$\chi^2_\mathcal{P}$ of this reweighed fit is flat and $\as$ is
determined based on the agreement with $\mathcal{D}$ (but based on
PDFs that have been modified to accommodate $\mathcal{P}$ at all
values of $\as$). If $\mathcal{P}$ determines $\as$ by itself (in the
sense of Sec~\ref{sec:thatprocessonly}) then the partial $\chi^2$ will
not be flat and will be used to obtain $\as$. A suitable interpolating
procedure between these two situations could be obtained in the
NNPDF framework by
minimizing as a function of $\{\theta\}$ and $\as$
\begin{equation}
	\mathrm{ERF}
	 = \chi^{2}\left[\{\theta\},\as,\mathcal{D}\right]
	+ w\chi^{2}\left[\{\theta\},\as,\mathcal{P}\right]
	\ ,
\end{equation}
where $w$ is a large number. Because of the cross validation based
regularization, the effect of $w$ will saturate either when we reach
$\chi_\mathcal{P}^2/d.o.f\approx 1$, so that only the first therm
varies as a function of $\as$, or else, if $\mathcal{P}$ determines
$\as$, the curvature of profile will exclusively depend on the second
term.

\section{Conclusions}
We argue that the uncertainties from determinations of $\as(M_Z)$
from hadronic data can be significatively reduced by interpreting them
as corrections on PDF based determinations, equivalent to adding the
data to the PDF fit. We propose a definition
a \emph{preferred value} of a parameter that may be advantageous when
studying theory uncertainties.

\subsection*{Acknwoledgments}

The work of Z.K. is supported by the European Research Council
Consolidator Grant ``NNLOforLHC2''.

\bibliographystyle{JHEP}
\bibliography{references}
\end{document}